\documentclass[aps,prd,twocolumn,groupedaddress]{revtex4}

\usepackage{graphicx}
\usepackage{amsmath}

\begin{document}

\title{Rotating optical cavity experiment testing Lorentz invariance at the \(10^{-17}\) level}

\author{S. Herrmann$^{1,2}$, A. Senger$^{1}$, K. M\"ohle$^{1}$, M. Nagel$^{1}$, E. V. Kovalchuk$^{1}$, A. Peters$^{1}$}

\affiliation{\sl $^{1}$ Institut f\"ur Physik, Humboldt-Universit\"at zu Berlin, Hausvogteiplatz 5-7, 10117 Berlin \\
$^{2}$ ZARM, Universit\"at Bremen, Am Fallturm 1, 28359 Bremen}

\date{\today}

\begin{abstract}
We present an improved laboratory test of Lorentz invariance in electrodynamics by testing the isotropy of the speed of light. Our measurement compares the resonance frequencies of two orthogonal optical resonators that are implemented in a single block of fused silica and are rotated continuously on a precision air bearing turntable. An analysis of data recorded over the course of one year sets a limit on an anisotropy of the speed of light of  $\Delta c/c \sim 1\times 10^{-17}$. This constitutes the most accurate laboratory test of the isotropy of $c$ to date and allows to constrain parameters of a Lorentz violating extension of the standard model of particle physics down to a level of $10^{-17}$.
\end{abstract}

\maketitle

The theory of special relativity formulated in 1905 \cite{Ein05} revealed Lorentz invariance as the universal symmetry of local space-time, rather than a symmetry of Maxwell's equations in electrodynamics alone. This striking insight was drawn from two postulates: (i) the speed of light in vacuum is the same for all observers independent of their state of motion, and (ii) the laws of physics are the same in any inertial reference frame. Today, local Lorentz invariance constitutes an integral part of the standard model of particle physics, as well as the standard theory of gravity, general relativity. Still, there have been claims that a violation of Lorentz invariance might arise within a yet to be formulated theory of quantum gravity \cite{KS89,AMU02,GP99a,GP99b,EMN99,CHK01}. Given a lack of quantitative predictions, the hope is to reveal a tiny signature of such a violation by pushing test experiments for Lorentz invariance across the board. An overview of recent such experiments can be found in \cite{Mat05}.

Previous measurements testing the isotropy of the speed of light, often referred to as modern Michelson-Morley experiments \cite{MM87}, have compared the resonance frequencies of optical \cite{MHB03,HSK05,HSK06,AOG05} or microwave \cite{WBC04,STW06} cavities, which were either actively rotated on a turntable or relied solely on Earth's rotation. The most precise of these have tested the isotropy of $c$ at an accuracy of a few parts in $10^{16}$ limited by relative resonator frequency stability.

\begin{figure}[h]
\includegraphics[width=0.48 \textwidth]{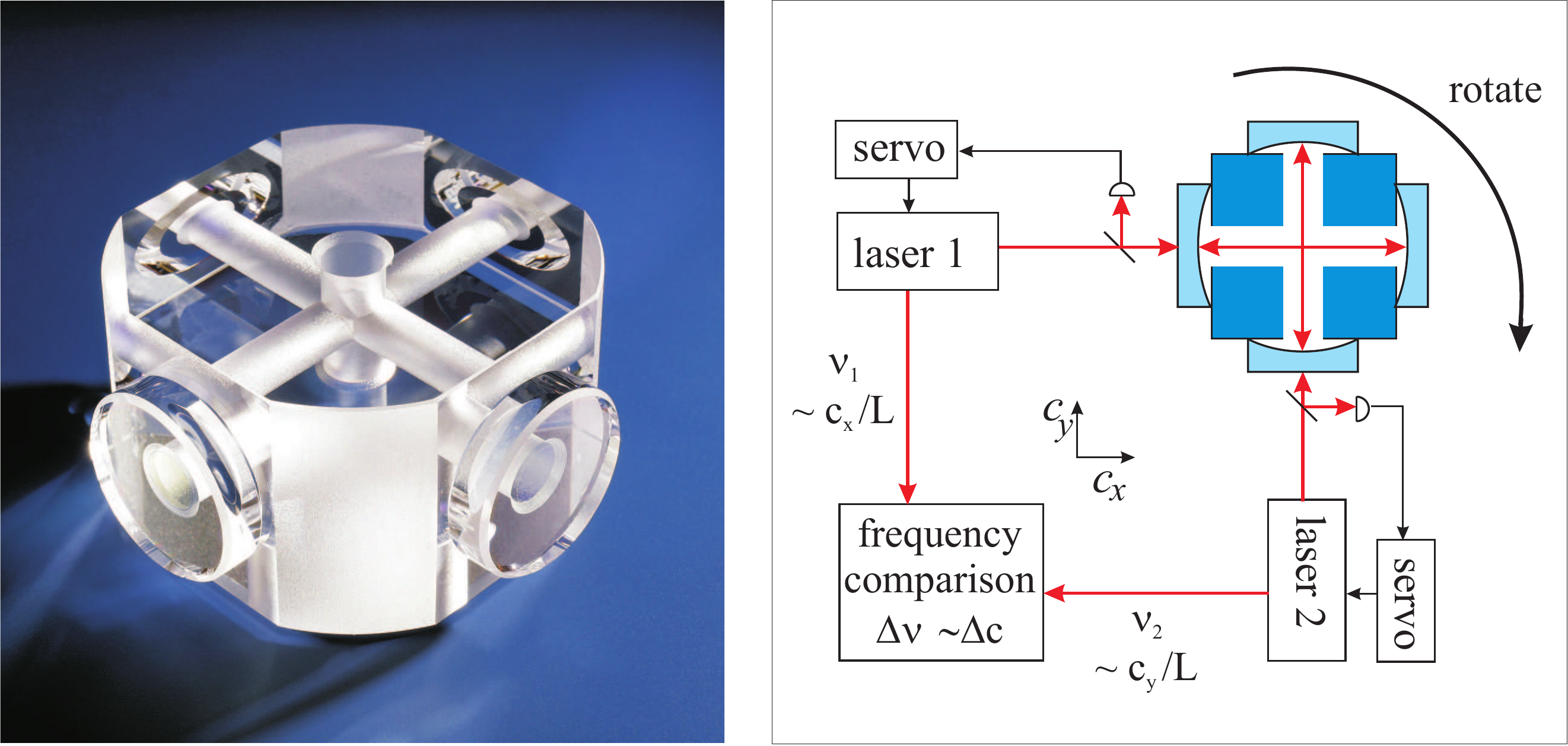}
\caption{Left: High-finesse fused silica resonators used in this experiment. Right: Basic principle of the experiment. The frequencies of two lasers, each stabilized to one of two orthogonal cavities, are compared during active rotation of the setup. (Photograph by E. Fesseler)}\label{fig1}
\end{figure}

The experiment presented here improves on this by one order of magnitude, based on an optimized cavity design and rotation on a precision turntable that allows to minimize systematic effects. The basic principle is depicted in Figure \ref{fig1}. At the core of the experiment are two crossed optical Fabry-P\'{e}rot resonators. We compare their resonance frequencies by stabilizing two Nd:YAG lasers to these cavities and taking a beat note measurement. The resonance frequency $\nu$ of a linear Fabry-P\'{e}rot cavity depends on the speed of light $c$ along its optical axes as given by
\begin{equation}
\nu = mc/2L
\end{equation}
where $m$ is an integer number and $L$ is the length of the resonator. Thus, to detect an anisotropy of the speed of light $\Delta c =  c_x - c_y$ we continuously rotate the setup and look for a modulation of the beat frequency $\Delta \nu$. Since the light in the cavities travels in both directions and $c$ refers to the two-way speed of light, such an isotropy violation indicating modulation would occur at twice the rotation rate.

\section{The experiment}\label{sec1}

The experiment applies a pair of crossed optical high-finesse resonators implemented in a single block of fused silica (Figure \ref{fig1}). This spacer block is a 55\,mm $\times$ 55\,mm $\times$ 35\,mm cuboid with centered perpendicular bore holes of 10\,mm diameter along each axis. Four fused silica mirror substrates coated with a high-reflectivity dielectric coating at $\lambda =  1064$\,nm are optically contacted to either side. The length of these two crossed optical resonators is matched to better than 2\,$\mu m$. The finesse of each resonator (TEM$_{00}$ mode) is $380\,000$, resulting in a linewidth of 7\,kHz. Two Nd:YAG lasers at $\lambda = 1064$\,nm are stabilized to these resonators using a modified Pound-Drever-Hall method \cite{DHK83}. Tuning and modulation of the laser frequency is achieved with piezo electric actuators attached to the laser crystal. Mechanical resonances of the piezo-electric actuators at $f_m = 444$\,kHz and 687\,kHz respectively are used for modulation of the laser frequencies. The light reflected from the cavities is detected and demodulated at $3 f_m$ to generate an error signal. Thermal effects from dissipation of laser power inside the resonators are minimized by coupling less than $50\,\mu$W optical power into the cavities.

\begin{figure}
\includegraphics[width=0.48\textwidth]{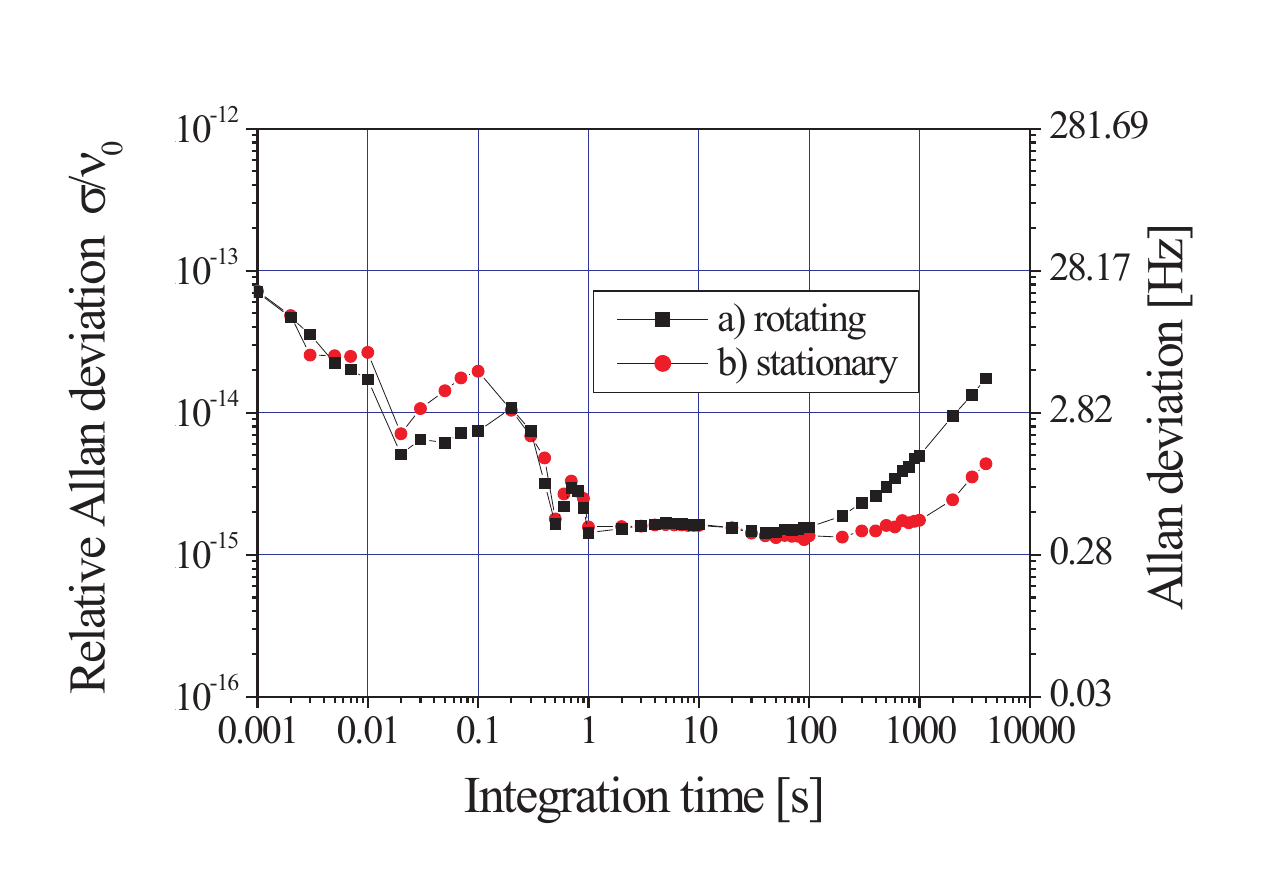}
\caption{Allan deviation calculated from a comparison of the two stabilized laser frequencies (a) while the setup is rotating and (b) with a stationary non rotating setup. \label{RAV}}
\end{figure}

\begin{figure}
\includegraphics[width = 0.4\textwidth]{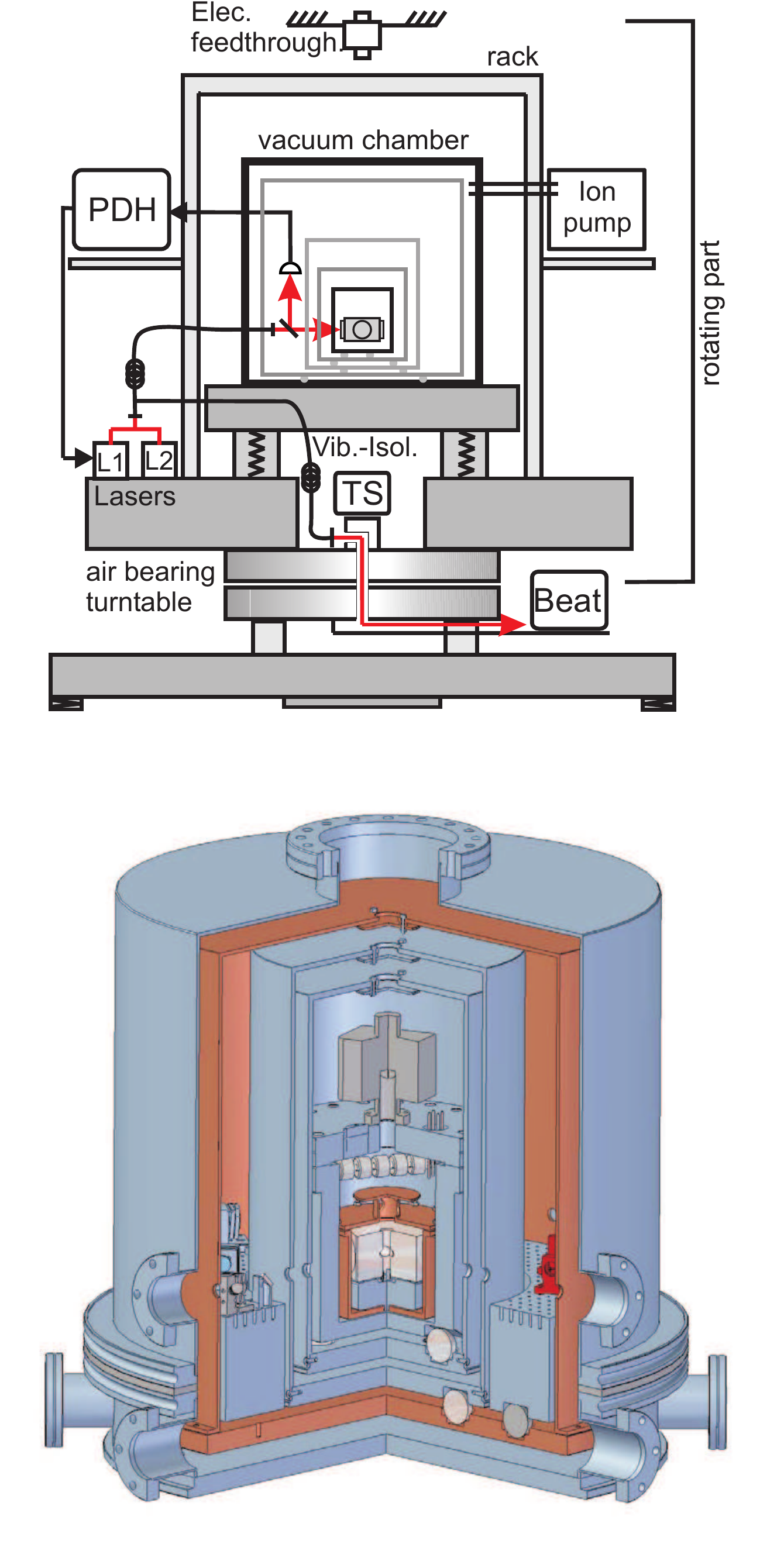}
\caption{Schematic of the complete rotating setup (top) and the custom-made vacuum chamber (bottom). TS=Tilt Sensor, PDH=Pound-Drever-Hall laser stabilization electronics.} \label{kammer}
\end{figure}

The cavities are installed in a vacuum chamber to ensure light propagation in vacuum inside the resonators and to reduce the influence of environmental noise, e.g.\ thermal fluctuations and vibrations in the laboratory. By placing the coupling optics inside the vacuum chamber as well, a high pointing stability of the laser beams incident on the resonators could be achieved. The custom-made vacuum chamber features several stages of thermal insulation and is placed upon an active vibration isolation system (\emph{HWL}, 350-M) (Figure \ref{kammer}).

To compare the stabilized laser frequencies, fractions of 1\,mW of each laser beam are split off and are overlapped on a fast photodiode to generate a beat note at the difference frequency $\Delta \nu = \nu_1 - \nu_2$. By choosing appropriate longitudinal modes of the cavities, this frequency is set to $< 2$\,GHz and counted with a sampling time interval of one second. Since the length of the two cavities is defined by a single monolithic block, drifts due to thermal expansion are largely the same for both resonators and thus cancel in a measurement of the difference frequency. We have observed a reduction of the relative drift to below 0.01\,Hz/s as compared to 100\,Hz/s absolute frequency drift of the individual resonators.

The stability of the frequency beat note is then characterized by calculating the Allan deviation. There is a pronounced flicker floor from 1\,s to 200\,s at a level of $\sim 1.5\times 10^{-15}$ (Figure \ref{RAV}) presumably caused by thermal noise of the mirror substrates.  This agrees with an estimate of the thermal noise level of our cavity based on a model of Numata {\sl et al.}\ \cite{NKC04}.

To enable continuous rotation of the resonators we employ an air bearing turntable which carries the complete laser stabilization setup (Figure \ref{kammer}) and rotates at a chosen rate of $T_{\rm tt} = 45$\,s. As opposed to using Earth´s rotation alone, such active rotation allows us to perform hundreds of rotations per day, while taking advantage of the excellent mid-term frequency stability of the cavities (Figure \ref{RAV}). On the other hand, active rotation potentially causes a systematic modulation of the beat frequency and might thus mimic an anisotropy signal. For example, gravitational or centrifugal forces that act on the resonators may get modulated with the turntable rotation and therefore modulate the length of the resonators. However, most of these effects lead to a modulation at a rate of $\omega_{\rm tt} = 2\pi/T_{\rm tt}$ so that they are in principle distinguishable from the anisotropy signal searched for at $2\omega_{\rm tt}$. Moreover, if the data spans more than one day, systematic effects with a fixed phase in the laboratory average out in the analysis for an anisotropy of $c$ that is fixed relative to a sidereal frame. Although such an analysis helps to discriminate a sidereal anisotropy signal from systematics, a large effort was still made to reduce systematic effects both at $2\omega_{\rm tt}$ and $\omega_{\rm tt}$.

First of all, we use a high precision air bearing turntable specified for $< 1 \mu$rad rotation axis wobble. Furthermore, we also prevent long-term variations of the rotation axis tilt, caused for example by daily fluctuations of the building tilt of several $\mu$rad. For this we apply an active stabilization \cite{HSK06} that keeps the rotation axis vertical to better than 1\,$\mu$rad, which reduces the effect from a periodic deformation of the cavities to frequency variations of less than 0.1\,Hz in amplitude. Effects from varying centrifugal forces are also reduced below an amplitude of 0.1\,Hz by an active stabilization of the rotation rate. Further measures include balancing the center of mass of the table ($< 1$\,mm offset from the rotation axis) and shielding the lasers and optics outside the vacuum chamber against air currents and temperature gradients in the laboratory.

At the chosen rotation period of 45\,s these measures reduce residual systematic frequency variations at $2\omega_{\rm tt}$ to amplitudes below 0.1\,Hz. This corresponds to a fractional frequency shift of $\Delta \nu/\nu_0 = 3 \times 10^{-16}$, which is well below the relative frequency stability of the beat note on the timescale of a single rotation (see Figure \ref{RAV}). While even faster rotation would have allowed to acquire more data and thus improve statistics, it resulted in increased residual systematic effects presumably due to modulated centrifugal forces and was thus not implemented.

Measurements with this setup have been performed intermittently during a time span of more than one year from May 2007 to June 2008. The total data includes recordings of the beat frequency, time and rotation angle at a sampling interval of 1\,s from more than $130\,000$ turntable rotations.

\section{Analysis for an anisotropy signal}

In what follows we first give a phenomenological, i.e.\ largely model-independent, description of an anisotropy signal and present results from a corresponding analysis. In Section \ref{sec2a} and \ref{sec2b} we then use these results to determine parameters of two different test theories for Lorentz violation. Throughout this analysis, we adopt the inertial Sun centered celestial equatorial coordinate frame (SCCEF) as used for the analysis of similar, previous such experiments \cite{KM02}. This coordinate system has the Z axis pointing north, the X axis pointing in the direction of the vernal equinox point, and the Y axis such that (X,Z,Y) form a right handed set.

Let us first consider the special case of the apparatus located at the North pole with the turntable rotation axis aligned with Earth's rotation axis. An anisotropy in the equatorial XY-plane ($c_X \neq c_Y$) then causes a modulation of the beat frequency with the rotation of the setup.  As noted above, this modulation would be at twice the rotation frequency, i.e. $2\omega_{\rm rot}$. If we fix the time axis relative to some arbitrary instant $t = 0$, we can describe this signal as
    \begin{equation}
        \frac{\nu_1 - \nu_2}{\nu_0} = \frac{\Delta \nu}{\nu_0} =  S' \sin 2\omega_{\rm rot}t + C' \cos 2\omega_{\rm rot}t, \label{signal}
    \end{equation}
where $\nu_0 \approx 282$\,THz and $S', C' \sim \frac{c_X - c_Y}{c}$. The rotation of the setup within the sidereal frame of reference is a superposition of turntable rotation $\omega_{\rm tt}$ and Earth's sidereal rotation at $\omega_{\oplus}$ such that  $\omega_{\rm rot} = \omega_{\oplus} \pm \omega_{\rm tt}$, plus or minus depending on the sense of turntable rotation. Since in our experiment we have $\omega_{\rm tt} \gg \omega_{\oplus}$, we can describe the anisotropy signal as a fast modulation at $2\omega_{\rm tt}$
\begin{equation}
        \frac{\Delta \nu}{\nu_0} = S \sin 2\omega_{\rm tt}t + C \cos 2\omega_{\rm tt}t, \label{signal1}
\end{equation}
with amplitudes $S$ and $C$ that slowly vary with Earth's rotation as given by
\begin{align}
        S & =  - C' \sin 2\omega_{\oplus}t + S' \cos 2\omega_{\oplus}t, \\
         \nonumber  \\
        C & = S' \sin 2\omega_{\oplus}t + C' \cos 2\omega_{\oplus}t.
\end{align}
This daily modulation is essential to discriminate an anisotropy signal from constant or slowly varying systematic effects caused by active rotation as described in Section \ref{sec1}. Only systematic effects subjected themselves to a 23.93 h modulation would mimic such a sidereal anisotropy signal.

Next, we consider an experiment located at an arbitrary geographical latitude $\chi$ such that Earth's axis and the turntable rotation axis do not coincide anymore. While this reduces sensitivity to an anisotropy in the equatorial XY-plane, it additionally provides sensitivity to an anisotropy in the XZ and YZ-plane. Furthermore, a modulation at $\omega_{\oplus}$ in addition to that at $2\omega_{\oplus}$ will appear. We thus generalize the above expressions to the following anisotropy signal (see \cite{KM02} for a formal derivation)
\begin{align}
        S & =  S_0 + S_{\rm s1} \sin (\omega_{\oplus}(t-t_0)) + S_{\rm c1}\cos (\omega_{\oplus}(t-t_0))  \nonumber \\
        & + S_{\rm s2} \sin (2\omega_{\oplus}(t-t_0)) + S_{\rm c2}\cos (2\omega_{\oplus}(t-t_0)), \label{signal2} \\
        \nonumber \\
        C & =  C_0 + C_{\rm s1} \sin (\omega_{\oplus}(t-t_0)) + C_{\rm c1}\cos (\omega_{\oplus}(t-t_0)) \nonumber  \\
        & + C_{\rm s2} \sin (2\omega_{\oplus}(t-t_0)) + C_{\rm c2}\cos (2\omega_{\oplus}(t-t_0))\:, \label{signal3}
\end{align}
where the phase is fixed by $t_0$ chosen in accordance to the adopted reference frame conventions. Again, this daily modulation at $\omega_{\oplus}$ and $2\omega_{\oplus}$ makes it possible to distinguish between a sidereal anisotropy of $c$ and systematic effects due to the active rotation.

To analyze our data for a Lorentz violation signal of the above form we proceed in two steps. First we determine the modulation amplitudes $C$ and $S$ as modeled in Eq.(\ref{signal1}) from short samples of the data set spanning 10 table rotations each. These samples extend over 450\,s each such that we may neglect a possible modulation due to Earth's rotation within each sample. We use a fit function of the form
\begin{align}
        \Delta \nu/ \nu &= S \sin (2\omega_{\rm tt} (t-t_0^\prime)) + C \cos (2\omega_{\rm tt} (t-t_0^\prime))\nonumber\\
        & \quad + A_S \sin (\omega_{\rm tt} (t-t_0^\prime)) + A_C \cos (\omega_{\rm tt} (t-t_0^\prime))\nonumber\\
        & \quad + A_0 + A_1 t \:,
\end{align}
where $A_0$ and $A_1$ account for an arbitrary offset and a linear drift while $A_S$ and $A_C$ account for residual systematics at $\omega_{\rm tt}$. In accordance with the reference frame convention of \cite{KM02} the starting time $t_0^\prime$ is determined by the first instant of the measurement at which one of the two resonators is oriented along the North-South direction. We choose a sample size of 10 rotations for each fit rather than fitting single rotations to reduce the correlation of a small linear drift and a sinusoidal variation of the beat frequency. We found, however, that choosing different sample sizes of $n = 2$ to 20 does not significantly change the final results.

\begin{figure*}
\includegraphics[width = 0.7\textwidth]{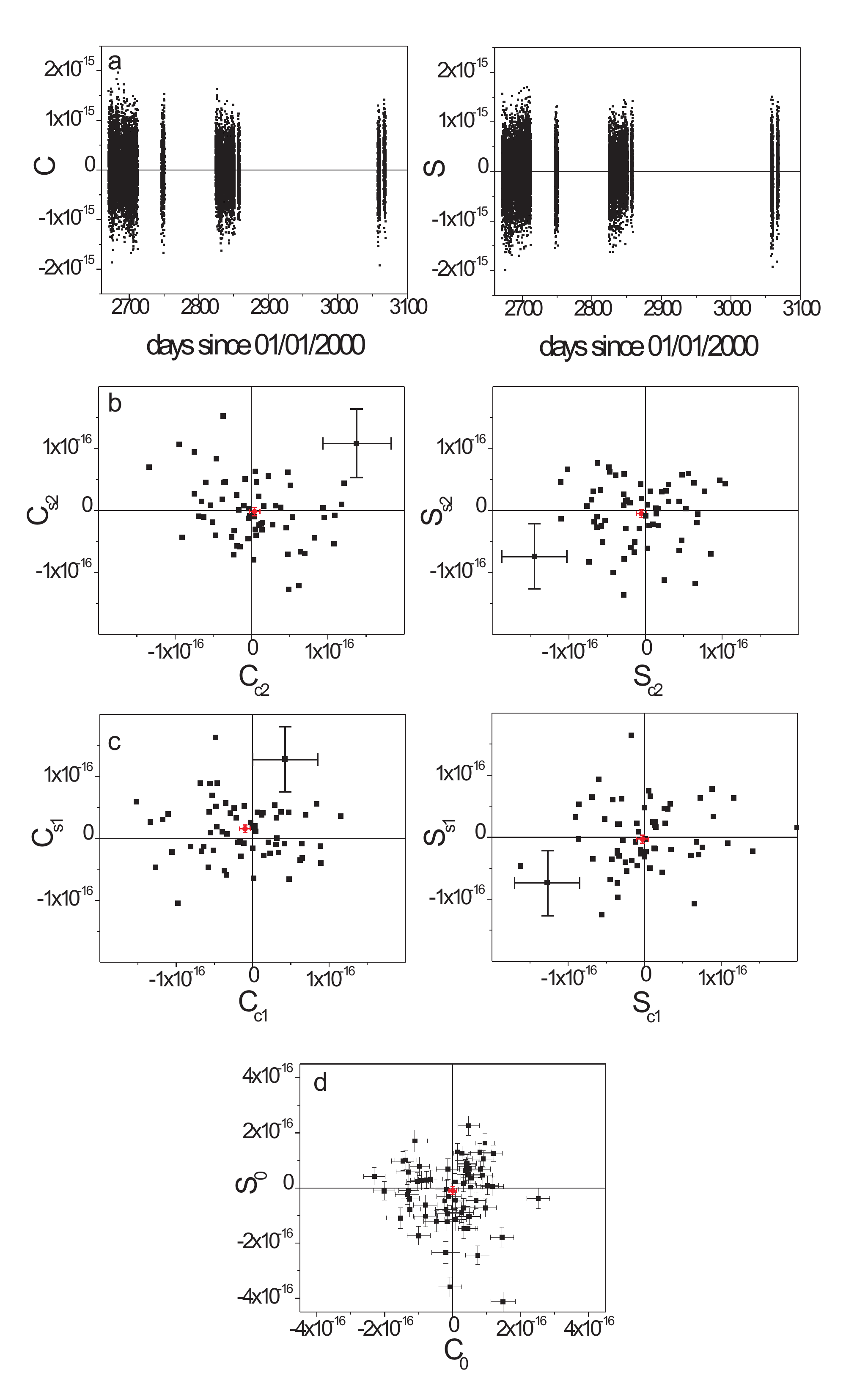}
\caption{{\bf Results I.} a): Cosine amplitudes $C$ (left) and Sine amplitudes $S$ (right) of a systematic beat frequency modulation at $2\omega_{\rm tt}$ for all $n = 13384$ measurement subsets (10 rotations each). $t_0^\prime$ is set to an instant when one of the resonators is oriented along the North South axis. b)+c): Amplitudes of a superimposed $2\omega_{\oplus}$  (11.96\,h) (b) respectively $\omega_{\oplus}$ (23.93\,h) (c) modulation of $C$ and $S$ respectively, as expected for an anisotropy of $c$ fixed within a sidereal frame. Each point represents the amplitudes determined from a 23.93\,h set of data. 64 such data sets are included. $t_0$ is set to an instant when the East-West axis of the laboratory coincides with the Y-axis of the adopted SCCEF reference frame. Error bars are omitted for the purpose of clarity except for one representative data point. d): Amplitudes $C_0$ and $S_0$ as modeled in equations (\ref{signal2}) and (\ref{signal3}), which are most prone to constant systematic effects (note the different scale). The mean values and standard errors (shown in red in (b),(c) and (d)) of the modulation amplitudes as modeled in equations (\ref{signal2}) and (\ref{signal3}) are: $C_0 = -0.2 \pm 11.7$, $C_{s1} = 15.5 \pm 6.0$, $C_{c1} = -9,9 \pm 7.2$, $C_{s2} = -1.5 \pm 6.6$, $C_{c2} = 4.0 \pm 6.6$ and $S_0 = -10.2 \pm 14.4$, $S_{s1} = -3.1 \pm 6.4$, $S_{c1} = -2,7 \pm 8.1$, $S_{s2} = -5.0 \pm 6.2$, $S_{c2} = -5.5 \pm 6.7$ (all values $\times 10^{-18}$). \label{results}}
\end{figure*}

From each sidereal day (23.93\,h) of measurement we obtain a distribution of 192 values of $S$ and $C$, and each value is assigned the mean time of the respective data sample. In total we obtain a distribution of 13384 values for $S$ and $C$ as shown in Figure \ref{results}a.

\begin{figure*}
\includegraphics[width = 0.7\textwidth]{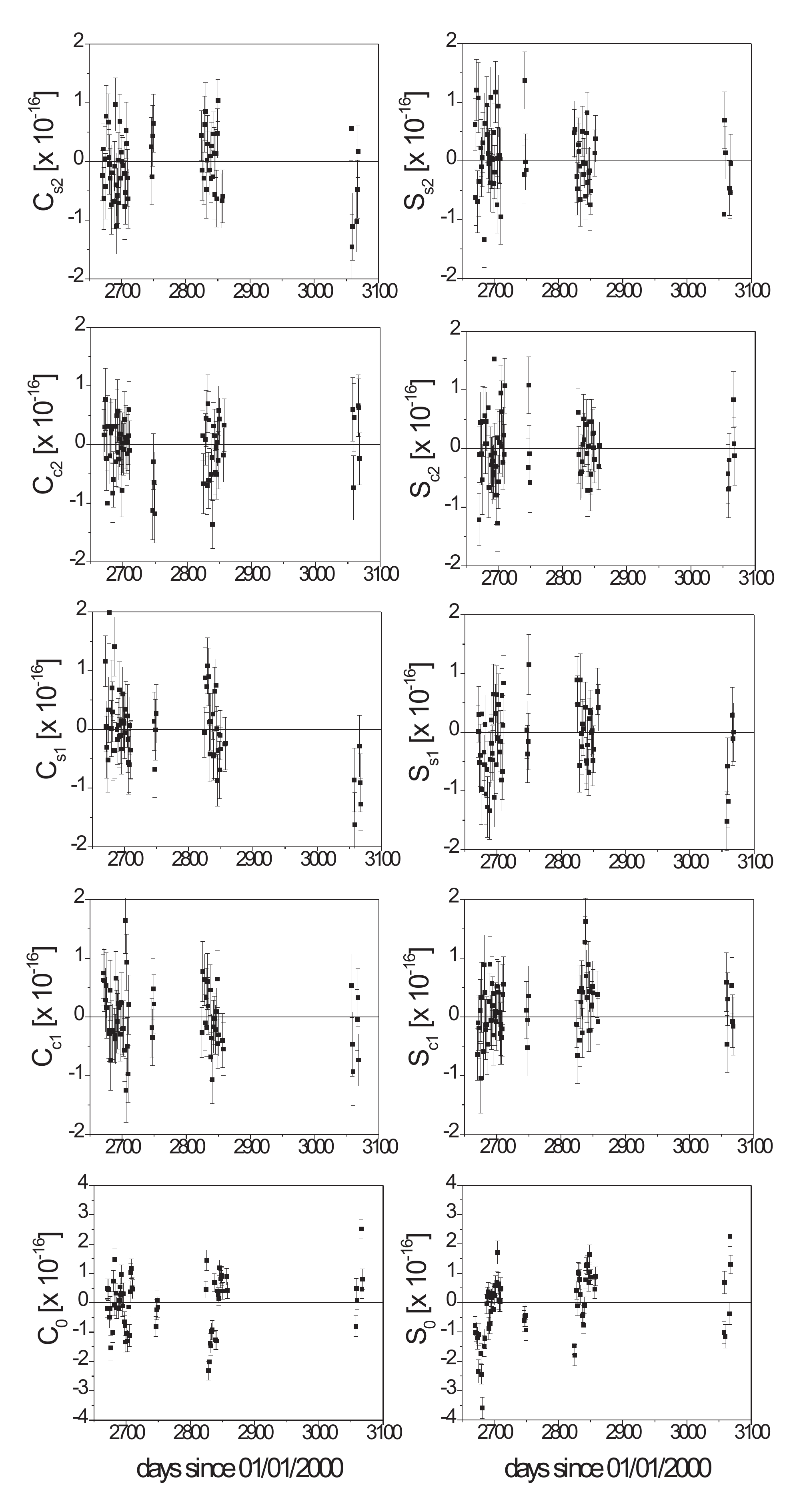}
\caption{{\bf Results II.} Data of Figure \ref{results} plotted vs time. As in Figure \ref{results} each point represents the amplitude determined from a 23.93\,h set of data.  See caption of Figure \ref{results} for more details. \label{results2}}
\end{figure*}

Next, each 23.93\,h interval of these distributions is fitted with equations (\ref{signal2}) and (\ref{signal3}) to determine whether there is any daily modulation as a consequence of a sidereal anisotropy of $c$. The results are shown in Figure \ref{results}b+c as well as in Figure \ref{results2}. Each graph corresponds to a pair of sidereal modulation amplitudes and shows a distribution of 64 data points. Each point is determined from one day of the measurement. The standard error associated with each one-day data point is on the order of $\sim 5 \times 10^{-17}$. If for each sidereal modulation amplitude we take the mean value of the corresponding 64 data points, we find a standard error on the order of  $\sim 7 \times 10^{-18}$ for each distribution and no deviation from zero by more than three standard errors.

Figure \ref{results}d shows the results for the modulation amplitudes $S_0$ and $C_0$. These are solely connected to a modulation at $2\omega_{\rm tt}$ (Eq.\eqref{signal1}, \eqref{signal2} and \eqref{signal3}) and thus are strongly effected by any residual systematic effects fixed to the laboratory frame. Single points of these amplitudes deviate by several standard errors from zero, however, over the complete measurement span of one year, the data points vary in magnitude and phase and thus average out. The variation of these amplitudes over time can also be seen from the bottom graphs in Figure \ref{results2}.

Overall, we conclude that no significant evidence for an anisotropy of $c$ fixed relative to a sidereal frame can be claimed from this data. This of course assumes otherwise uncorrelated noise, e.g.\ no annual phase shift of a non-zero anisotropy signal.

\subsection{Analysis in the framework of the minimal standard model extension}\label{sec2a}

The above phenomenological results can be further evaluated as a test of Lorentz invariance in electrodynamics, adopting the Lorentz violating extension of the standard model of particle physics  by D. Colladay and V.A. Kosteleck\'{y} {\sl et al.}\ \cite{CK97,CK98}. In this standard model extension (SME), Lorentz violation in electrodynamics is modeled by extending the Lagrangian of the photonic sector $\mathcal{L} = - \frac{1}{4} F^{\mu\nu}F_{\mu\nu}$  with a term  $\mathcal{L}_{\rm ext} = - \frac{1}{4}(k_F)_{\mu\nu\kappa\lambda} F^{\mu\nu}F^{\kappa\lambda}$ where $F_{\mu\nu}$ is the electrodynamic field tensor and $(k_F)\mu\nu\kappa\lambda$  a tensor that parameterizes Lorentz violation with 19 independent components.

V.A. Kosteleck\'{y} and M. Mewes \cite{KM02}  have shown that with such an extra term the propagation of light in vacuum can be described in analogy to the propagation of light in an anisotropic medium. They have also modeled how this anisotropy affects the resonance frequency of a linear optical Fabry-P\'{e}rot resonator. These results can be used to model the amplitude coefficients of equations (\ref{signal2}) and (\ref{signal3}) for our experiment.

The resulting expressions as derived explicitly in \cite{HSK06} are given in Table \ref{SME-coefficients}. Ten of the SME parameters are linked to birefringence and are restricted to values $< 10^{-32}$  by astrophysical measurements \cite{KM02,Alt06}. These parameters are assumed to be zero here.  Eight of the remaining nine SME parameters, grouped into two traceless 3x3 matrices $\kappa_{e-}$ and $\kappa_{o+}$, can then be determined from the present measurement: Five are parity even and boost independent ($\kappa_{e-}$, symmetric) and three are parity odd and boost dependent ($\kappa_{o+}$, antisymmetric). The boost dependent parameters $\kappa_{o+}$ lead to an annual phase shift of the anisotropy signal due to Earth's orbital revolution (Table \ref{SME-coefficients}). Note that since our measurement spans more than one year we are indeed able to resolve such an annual variation.

A simultaneous fit of the expressions in Table \ref{SME-coefficients} to the $2 \times 5$ distributions of sidereal modulation amplitudes obtained from our data, yields estimates on the eight SME parameters as summarized in Table \ref{Tab1}. Four parameters of $\kappa_{e-}$ and the three boost dependent parameters of $\beta_\oplus\kappa_{o+}$ (with Earth's orbital boost $\beta_\oplus=v_{\oplus}/c = 10^{-4}$) feature a standard error of $\sim 1 \times 10^{-17}$, while one parameter, $\kappa_{e-}^{ZZ}$, shows a slightly increased error bar of $1.7 \times 10^{-17}$. This is attributed to the fact that $\kappa_{e-}^{ZZ}$ enters the $C_0$ component only, which is most prone to the residual systematic effects as discussed above in Section \ref{sec2a}.

All together these limits represent a significant improvement of more than one order of magnitude over the results of the best previous experiment by Stanwix {\sl et al.}\ \cite{STW06}. They also complement results from a newly emerging astrophysical technique, which argues that equivalent limits at the $10^{-18}$ level can be obtained by analyzing observations of ultra-high energy cosmic rays \cite{KR08}.

\begin{table}
\caption{Estimates on photonic SME parameters obtained from this work (one sigma errors). For comparison the limits obtained by Stanwix {\sl et al.}\ \cite{STW06} are also given. All values are $\times 10^{-17}$. $\beta_{\oplus} = v_{\oplus}/c = 10^{-4}$ accounts for Earth's
orbital boost. \label{Tab1}}{
\begin{tabular}{c|cc}
\hline\hline \rule{0pt}{12pt}
& this work & Stanwix {\sl et al.}\ \cite{STW06} \\ \hline
$\kappa_{e-}^{XY}$ & -0.31 $\pm$ 0.73 & 29 $\pm$ 23 \\
$\kappa_{e-}^{XZ}$ & 0.54 $\pm$ 0.70 & -69 $\pm$ 22 \\
$\kappa_{e-}^{YZ}$ & -0.97 $\pm$ 0.74 & 21 $\pm$ 21 \\
$\kappa_{e-}^{XX} - \kappa_{e-}^{YY}$ & 0.80 $\pm$ 1.27 & -50 $\pm$ 47 \\
$\kappa_{e-}^{ZZ}$ & -0.04 $\pm$ 1.73 & 1430 $\pm$ 1790 \\
$\beta_{\oplus}\kappa_{o+}^{XY}$  &-0.14 $\pm$ 0.78 & -9 $\pm$ 26 \\
$\beta_{\oplus}\kappa_{o+}^{XZ}$ & -0.45 $\pm$ 0.62 & -44 $\pm$ 25 \\
$\beta_{\oplus}\kappa_{o+}^{YZ}$ &  -0.34 $\pm$ 0.61 & - 32 $\pm$ 23 \\
\hline\hline
\end{tabular}}
\end{table}

\begin{table*}
\centering
\begin{tabular}{l c}
\hline\hline\noalign{\smallskip}   & SME amplitude \\
\noalign{\smallskip} \hline \noalign{\smallskip}
$C_{\textnormal{0}}$: & $\gamma_0 \left(\frac{3}{2}\tilde\kappa_{e-}^{ZZ} - \beta_{\oplus}[(\cos\eta\tilde\kappa_{o+}^{XZ} + 2 \sin\eta\tilde\kappa_{o+}^{XY})\cos\Omega_{\oplus}T + \tilde\kappa_{o+}^{YZ}\sin\Omega_{\oplus}T] \right)$  \\
%&\\
$C_{\textnormal{s1}}$: & $\gamma_1 \left( -\tilde\kappa_{e-}^{YZ} + \beta_{\oplus}[\cos\eta\tilde\kappa_{o+}^{XY} - \sin\eta\tilde\kappa_{o+}^{XZ}]\cos\Omega_{\oplus}T \right)$ \\
%&\\
$C_{\textnormal{c1}}$: & $\gamma_1 \left( -\tilde\kappa_{e-}^{XZ} + \beta_{\oplus}[\sin\eta\tilde\kappa_{o+}^{YZ}\cos\Omega_{\oplus}T - \tilde\kappa_{o+}^{XY}\sin\Omega_{\oplus}T] \right)$ \\
%&\\
$C_{\textnormal{s2}}$: & $\gamma_2 \left( \tilde\kappa_{e-}^{XY} - \beta_{\oplus}[\cos\eta\tilde\kappa_{o+}^{YZ}\cos\Omega_{\oplus}T + \tilde\kappa_{o+}^{XZ}\sin\Omega_{\oplus}T] \right)$ \\
%&\\
$C_{\textnormal{c2}}$: &  $\gamma_2 \left( \frac{1}{2}[\tilde\kappa_{e-}^{XX} - \tilde\kappa_{e-}^{YY}] - \beta_{\oplus}[\cos\eta\tilde\kappa_{o+}^{XZ}\cos\Omega_{\oplus}T - \tilde\kappa_{o+}^{YZ}\sin\Omega_{\oplus}T] \right)$ \\
\noalign{\smallskip}\hline\noalign{\smallskip}
$S_{\textnormal{0}}$: & 0 \\
%&\\
$S_{\textnormal{s1}}$: & $\frac{\gamma_3}{\gamma_1} C_{\textnormal{c1}}$ \\
%&\\
$S_{\textnormal{c1}}$: & $-\frac{\gamma_3}{\gamma_1} C_{\textnormal{s1}}$ \\
%&\\
$S_{\textnormal{s2}}$: & $-\frac{\gamma_4}{\gamma_2} C_{\textnormal{c2}}$ \\
%&\\
$S_{\textnormal{c2}}$: & $\frac{\gamma_4}{\gamma_2} C_{\textnormal{s2}}$ \\
\noalign{\smallskip}\hline\hline
\end{tabular}
\caption[Modulation amplitudes related to photonic SME parameters]{\slshape Modulation amplitudes according to equations
(\ref{signal2}) and (\ref{signal3}) related to photonic SME
parameters. $\gamma_0 = \frac{1}{4}\sin^2\chi$, $\gamma_1 =
\frac{1}{2}\cos\chi\sin\chi$, $\gamma_2 = \frac{1}{4}(1 +
\cos^2\chi)$, $\gamma_3 = -\frac{1}{2}\sin\chi$ and $\gamma_4 = \frac{1}{2}\cos\chi$. Relations are stated to first order in orbital boost. $\beta_{\oplus} = 10^{-4}$ is the boost parameter, $\chi = 37^{\circ}$ is the colatitude of the Berlin laboratory and $\eta = 23^{\circ}$ is the tilt of Earth's axis relative to the SCCEF
$Z$-axis. In accordance to the reference frame conventions in \cite{KM02}. T = 0 is set to the instant of Earth passing vernal equinox. \label{SME-coefficients}}
\end{table*}

\subsection{Analysis in the Mansouri-Sexl framework}\label{sec2b}

We also analyze the data according to the kinematic test theory of R. Mansouri and R.U. Sexl \cite{MS76}, which builds on earlier work by H.P Robertson \cite{Rob46}. In this test theory a preferred frame is assumed in which the speed of light $c$ is isotropic, usually taken to be the cosmic microwave background. General, linear transformations, using three free parameters $\alpha, \beta, \delta$, transform from this preferred frame to a frame moving at a velocity $v$. In the moving frame an anisotropy of the propagation of light then takes the form $\Delta c/c = (\beta + \delta -\frac{1}{2}) v^2/c^2 \sin^2\theta$ where $\theta$ is the angle between the direction of the propagation of light and the direction of $v$. For $\alpha = \frac{1}{2}, \beta = \frac{1}{2}, \delta = 0$, the generalized transformations reduce to Lorentz transformations and no anisotropy of $c$ is observed.

A derivation of the signal amplitudes of equations (\ref{signal2}) and (\ref{signal3}) in the Mansouri-Sexl framework has been given in \cite{HSK06}. The resulting expressions are given in Table \ref{RMScoefficients}. Therein we take the velocity of the laboratory relative to the CMB as the superposition of the solar system's velocity $v_{c} = 370$\,km/s, pointing towards $\psi = 100^{\circ}$ right ascension and $\phi = -7^{\circ}$ declination and the annual modulation due to Earth's orbit with $v_{\oplus} = 30$\,km/s.

Simultaneously fitting these expressions to our data yields a value of $(\beta + \delta -\frac{1}{2}) = (4 \pm 8) \times 10^{-12}$. This is a factor of 10 more stringent as compared to the value of $(9.4 \pm 8.1) \times 10^{-11}$ given by Stanwix {\sl et al.}\ \cite{STW06}.

\begin{table*}
\centering
\begin{tabular}{l c}
\hline\hline\noalign{\smallskip}  & RMS amplitude $\left(\times \left(\beta + \delta -\frac{1}{2}\right) \frac{v_c^2}{c^2}\right)$ \\
\noalign{\smallskip} \hline  \noalign{\smallskip}
$C_{\textnormal{0}}$: &  $\frac{1}{2}\gamma_0(-1 + 3\cos2\phi) + 2\frac{v_{\oplus}}{v_c}\gamma_0(\sin\psi\cos\phi\cos\eta - 2\sin\phi\sin\eta)\cos\Omega_{\oplus}T + 2\frac{v_{\oplus}}{v_c}\gamma_0\cos\phi\cos\psi\sin\Omega_{\oplus}T$\\
%& \\
$C_{\textnormal{s1}}$:  &  $-\gamma_1\sin\psi\sin2\phi - 2\frac{v_{\oplus}}{v_c}\gamma_1(\sin\phi\cos\eta + \sin\psi\cos\phi\sin\eta)\cos\Omega_{\oplus}T$ \\
%& \\
$C_{\textnormal{c1}}$: &  $-\gamma_1 \cos\psi\sin2\phi - 2 \frac{v_{\oplus}}{v_c}\gamma_1\cos\psi\cos\phi\sin\eta\cos\Omega_{\oplus}T - 2 \frac{v_{\oplus}}{v_c}\gamma_1\sin\phi\sin\Omega_{\oplus}T $ \\
%& \\
$C_{\textnormal{s2}}$: & $-\gamma_2 \sin2\psi\cos^2\phi - 2\frac{v_{\oplus}}{v_c}\gamma_2\cos\psi\cos\phi\cos\eta\cos\Omega_{\oplus}T- 2\frac{v_{\oplus}}{v_c}\gamma_2\sin\psi\cos\phi\sin\Omega_{\oplus}T$\\
%& \\
$C_{\textnormal{c2}}$: &  $-\gamma_2 \cos2\psi\cos^2\phi + 2\frac{v_{\oplus}}{v_c}\gamma_2\sin\psi\cos\phi\cos\eta\cos\Omega_{\oplus}T- 2\frac{v_{\oplus}}{v_c}\gamma_2\cos\psi\cos\phi\sin\Omega_{\oplus}T$ \\
\noalign{\smallskip}\hline\noalign{\smallskip}
$S_{\textnormal{0}}$: & 0  \\
%& \\
$S_{\textnormal{s1}}$: & $\frac{\gamma_3}{\gamma_1}C_{\textnormal{c1}}$ \\
%& \\
$S_{\textnormal{c1}}$: & $-\frac{\gamma_3}{\gamma_1}C_{\textnormal{s1}}$  \\
%& \\
$S_{\textnormal{s2}}$: & $-\frac{\gamma_4}{\gamma_2}C_{\textnormal{c2}}$  \\
%& \\
$S_{\textnormal{c2}}$: & $\frac{\gamma_4}{\gamma_2}C_{\textnormal{s2}}$  \\
\noalign{\smallskip}\hline\hline
\end{tabular}
\caption[Modulation amplitudes related to RMS parameter]{\slshape
Modulation amplitudes according to equations (\ref{signal2}) and
(\ref{signal3}) related to the RMS parameter $(\beta + \delta -\frac{1}{2})$. $\gamma_0 =
\frac{1}{4}\sin^2\chi,\gamma_1 =\frac{1}{2}\sin\chi\cos\chi ,\gamma_2 =
\frac{1}{4}(1 +\cos^2\chi) ,\gamma_3 = -\frac{1}{2}\sin\chi$ and $\gamma_4 = \frac{1}{2}\cos\chi$. $\chi = 37^{\circ}$ denotes the laboratory colatitude, $\eta = 23^{\circ}$ the tilt of Earth's axis relative to the SCCEF $Z$-axis. $\psi = 100^{\circ}$ right ascension and $\phi = -7^{\circ}$ declination denote the direction of the solar system´s velocity relative to the cosmic microwave background. Terms varying with Earth's orbital motion are suppressed by $v_{\oplus}/v_c \sim0.08$. Also here T = 0 is set to the instant of Earth passing vernal equinox. \label{RMScoefficients}}
\end{table*}

\section{Conclusion}
In conclusion, we have set a limit on an anisotropy of the speed of light at a level of $\Delta c/c \sim 1\times 10^{-17}$, which allows us to confirm the validity of Lorentz invariance in electrodynamics at the $10^{-17}$ level. This accuracy has been obtained with optical resonators that feature a relative frequency stability of  $\Delta \nu/\nu_0 \sim 1 \times 10^{-15}$ in 1\,s. The final precision could be reached by integrating over more than $130\,000$ rotations relying on a careful suppression of systematic effects caused by the turntable rotation. \\
Finally, we note that comparable results from a similar experiment \cite{ENS09} have been reported after submission of this manuscript.

The relative frequency stability is currently limited by thermal noise of the cavity mirrors. Thus, in the longer term it should be possible to improve the relative frequency stability by using cryogenic resonators \cite{MHB03,Ben08}. Together with a reasonable improvement in the suppression of systematic effects, this would ultimately allow one to test for potential violations of Lorentz invariance in electrodynamics in the $\Delta c/c \sim 10^{-20}$ regime.

We thank G. Ertl for his support and H. M\"uller for valuable discussions. S. H. acknowledges support from the Studienstiftung des deutschen Volkes.

\end{document}